\documentclass[%
 preprint, showkeys,
 superscriptaddress,
 amsmath,amssymb,
 aps, physrev,
]{revtex4-2}

\usepackage{submission-author-macros}
\usepackage{graphicx}
\usepackage{dcolumn}
\usepackage{bm}

\begin{document}


\title{\textbf{Dynamical diffraction formalism for imaging time-dependent diffuse scattering from coherent phonons with Dark-Field X-ray Microscopy} 
}%

\author{Darshan Chalise}
\affiliation{Materials Science and Engineering, Stanford University, Stanford, CA 94305, USA}
\affiliation{SLAC National Accelerator Laboratory, Menlo Park, CA 94025, USA}
\affiliation{SIMES, Stanford University, Stanford, CA 94305, USA}

\author{Brinthan Kanesalingam}
\affiliation{Materials Science and Engineering, Stanford University, Stanford, CA 94305, USA}
\affiliation{SLAC National Accelerator Laboratory, Menlo Park, CA 94025, USA}
\affiliation{SIMES, Stanford University, Stanford, CA 94305, USA}
\affiliation{PULSE Institute, Stanford University, Stanford, CA 94305, USA}

\author{Dorian P. Luccioni}
\affiliation{Materials Science and Engineering, Stanford University, Stanford, CA 94305, USA}
\affiliation{SLAC National Accelerator Laboratory, Menlo Park, CA 94025, USA}
\affiliation{PULSE Institute, Stanford University, Stanford, CA 94305, USA}

\author{Daniel Schick}
\affiliation{Max-Born-Institut für Nichtlineare Optik und Kurzzeitspektroskopie, Berlin, Germany}

\author{Aaron M. Lindenberg}
\affiliation{Materials Science and Engineering, Stanford University, Stanford, CA 94305, USA}
\affiliation{SLAC National Accelerator Laboratory, Menlo Park, CA 94025, USA}
\affiliation{SIMES, Stanford University, Stanford, CA 94305, USA}
\affiliation{PULSE Institute, Stanford University, Stanford, CA 94305, USA}

\author{Leora Dresselhaus-Marais}
\CorrespondingAuthorEmail{leoradm@stanford.edu}
\affiliation{Materials Science and Engineering, Stanford University, Stanford, CA 94305, USA}
\affiliation{SLAC National Accelerator Laboratory, Menlo Park, CA 94025, USA}
\affiliation{SIMES, Stanford University, Stanford, CA 94305, USA}
\affiliation{PULSE Institute, Stanford University, Stanford, CA 94305, USA}


\begin{abstract}
Coherent acoustic phonons, whose damping sets the upper bound of quality factors in acoustic resonators, play a critical role in advanced telecommunication and quantum information technologies. Yet, probing their decay in the GHz regime remains challenging  using conventional surface-based techniques. Dark-field X-ray microscopy (DFXM) offers a solution by enabling through-depth, non-destructive and full-field imaging of strain fields and dislocations inside bulk materials with high spatial and angular resolution. We previously used kinematic diffraction theory to describe DFXM signals based on how the Bragg peak shifts due to the strain wave, allowing us to reconstruct the frequency spectrum of coherent phonons as a function of depth through the sample. The approach of tracking the Bragg peak shifts to study phonon dynamics, however, places an upper-bound to the highest phonon frequency that can be studied, determined by the spatial resolution of the measurement. In this work, we discuss how coherent phonon dynamics can be studied with DFXM from  time-dependent intensity oscillation sidebands. This approach simultaneously allows studying coherent phonon dynamics in real and reciprocal space, overcoming frequency resolution limits imposed by the real-space resolution of Bragg-peak tracking. Using Takagi-Taupin dynamical diffraction formalism, we establish the spatial and reciprocal space resolution achievable for studying the coherent phonon dynamics and evaluate conditions for observing long-lived intensity oscillations. We close by proposing experimental strategies to optimize excitation bandwidths and reciprocal-space selectivity. The formalism in the paper enables the design of DFXM experiments for quantitative, frequency-resolved measurements of acoustic phonon decay and phonon-defect interactions in bulk crystalline materials. 
\end{abstract}

\maketitle


\section{Introduction}

Coherent acoustic phonons are collective in-phase oscillations resulting in a macroscopic change in lattice spacing \cite{lindenberg2000time}. Measuring the damping of coherent phonons is vital for determining the upper bound of the quality factor $Q$ in GHz acoustic resonators \cite{chalise2025formalism}, which are essential for applications in telecommunication \cite{aigner20193g} and quantum information processing \cite{bienfait2019phonon}. Their decay, particularly in the GHz frequency regime, is difficult to probe due to the challenges of surface measurements with  picosecond acoustic experiments \cite{daly2009picosecond} with optical pump and probe beams. Quantifying and decoupling the contributions of coherent phonon damping from the Akhiezer mechanism \cite{daly2009picosecond, liao2018akhiezer}, three-phonon scattering processes \cite{daly2009picosecond, liao2018akhiezer} and subsurface defects and dislocations remains an open challenge that demands advanced experimental techniques \cite{chalise2025formalism}.

Dark-field X-ray microscopy (DFXM) is an imaging technique that enables full-field, real-space visualization of X-ray diffraction signals from crystals by placing an objective lens along the diffracted beam path \cite{simons2015dark, poulsen2021geometrical}. DFXM allows for non-destructive imaging of strain fields and defects deep within bulk materials with high spatial and angular resolution \cite{poulsen2021geometrical}. Experimentally, DFXM experiments achieve spatial resolution approaching 150 nm  using an objective lens, which simultaneously magnifies the image onto a detector and serves as an aperture in reciprocal space . When combined with monochromatic and nearly collimated X-rays from 4th generation synchrotrons or X-ray free-electron lasers (XFELs), DFXM can reach reciprocal space resolutions better than \(10^{-4}\) radians \cite{poulsen2021geometrical}. These capabilities make DFXM particularly powerful for imaging and quantifying localized strain fields associated with dislocations in three dimensions within bulk crystalline materials \cite{poulsen2021geometrical, kanesalingam2025computation}. With its ability to simultaneously resolve strain in real and reciprocal space over a large field of view, DFXM overcomes limitations of traditional time-resolved acoustic methods and enables studying acoustic decay mechanisms and phonon-defect interactions \cite{chalise2025formalism}.

In our previous work \cite{chalise2025formalism}, we used the kinematic theory of diffraction to understand how time-resolved strain measured from the shifts in Bragg peaks  could be used to study the propagation and decay of GHz strain waves.  However, work by Lindenberg \textit{et al.} \cite{lindenberg2000time} showed that X-ray scattering from coherent acoustic phonons also leads to time-dependent oscillations in scattering intensity away from the Bragg peak. These oscillations exhibit frequencies that correspond precisely to the phonon frequency, and the scattering occurs at wave vectors offset from the Bragg peak ($\vec{G}_{hkl}$) by the phonon wave vector
\begin{equation}
\vec{K}_{\text{scattering}} = \vec{K}_{\text{phonon}} \pm \vec{G}_{hkl}.
\end{equation}

Although the above equation can be derived purely from a kinematic wave-vector matching consideration, a purely kinematic diffraction with no absorption results in infinitesimally narrow Bragg peaks for large crystals with no overlap of peaks for two strain states. In such theoretical cases, no intensity oscillations would occur \cite{lindenberg2001}.  To explain the intensity oscillations at the sidebands for bulk crystals, Lindenberg \cite{lindenberg2001} employed the expanding crystal model and the Takagi–Taupin equations for X-ray scattering during strain wave propagation \cite{larson1980x}. These equations enable a rigorous treatment of dynamical diffraction in the presence of time-varying strain and demonstrate how, in the case of coherent acoustic phonons, the scattering wave vector and the intensity oscillations correspond directly to specific phonon wave vectors and frequencies \cite{lindenberg2000time}. 

In this paper, we employ the same dynamical diffraction formalism for imaging the time-dependent intensity oscillations of the X-ray scattering from coherent phonons using DFXM. The formalism of DFXM to study such phenomena allows for simultaneous mapping of phonon population in momentum space and their propagation in real space using time-resolved measurements. Importantly, as established in the previous work using kinematic theory, the spatial resolution of DFXM places an upper limit on the reconstructible frequency when performing frequency-resolved measurements from tracking the change in the position of the  Bragg peak. We overcome that limit when studying phonon frequencies from sidebands, where the limits, are purely determined by the reciprocal space resolution.

We first revisit the formalism for the Takagi–Taupin equations as applied to the coherent phonon wavepacket. We establish how individual frequencies and populations of coherent phonons can be mapped from diffuse scattering intensities. These intensities appear at specific momentum transfers corresponding to the phonon wavevector and share the same frequency as the oscillating strain in the wave. We then establish, from the Takagi–Taupin equations, the spatial resolution that can be achieved for acoustic wave propagation when imaged with DFXM. Next, we build the reciprocal space resolution for the case of phonon sidebands that result in time-dependent oscillations in intensities. Finally, using the established reciprocal space resolution, we determine the expected decay time of the time-dependent intensity oscillations for the sidebands from coherent phonons generated using excitation of metallic transducers. We further provide optimization strategies narrow the frequency bandwidth of the coherent phonon wavepacket. In addition, we discuss how to optimize the reciprocal space resolution of the measurement to enable long-lived intensity oscillations, which can be used for long-duration measurements to quantify acoustic decay. Finally, we provide a tool with parameters of transducer thickness, laser power and XFEL bandwidth and divergences to simulate DFXM images of the coherent acoustic phonon propagation and the corresponding intensity oscillations of the phonon-sidebands.

\section{The Takagi-Taupin implementation for coherent phonons}

The Takagi–Taupin equations enable computing the scattering amplitude of x-rays using the dynamical theory of diffraction. In the current work, we follow the formalism by Klar \& Rustichelli \cite{klar1973dynamical} to compute the x-ray scattering amplitude and thereby the x-ray diffraction intensity in the case of a crystal which has depth-dependent strain. This allows us to compute the rocking-curve of the diffraction as a function of time as the acoustic wave propagates. We note that the developed formalism is specific for imaging  for symmetric Bragg geometry of scattering which is described by Larson \& Barhorst \cite{larson1980x}. For the case of coherent acoustic phonons, this formalism is consistent with the observation that the intensity oscillations away from the center of the Bragg peaks follow an oscillation frequency 
\begin{equation}
\omega = v |G| \Delta \theta \cot \theta_B. 
\end{equation}
Here, \(v\) is the longitudnal velocity of sound in the material, \(|G|\) is the lattice vector of the plane of diffraction, \(\theta_B\) is the center of the Bragg peak, and \(\Delta \theta\) is the deviation away from the Bragg peak.

We begin this section with a brief review of the formalism by Klar \& Rustichelli \cite{klar1973dynamical}.  We then implement the formalism to obtain the results by Lindenberg \textit{et al.} \cite{lindenberg2000time}, establishing the relationship between \(\Delta \theta\) and \(\omega\). Finally, we show, the weight of a certain frequency \(\omega\) in a phonon spectrum is exactly reproduced by the weight of the intensity oscillation at the same frequency. 

In the formalism by Klar \& Rustichelli \cite{klar1973dynamical}, we define a length parameter given by  

\begin{equation}
    A=\frac{r_e\,f(\psi)\,\lambda\,x}{V_c\sin(\theta_B)},
\end{equation}

where, $r_e$ is the classical electron radius, $f(\psi) = f'(\psi)+if(\psi) $ is the structure factor at a scattering angle $\psi$, $\lambda$ is the wavelength of the x-ray used, $V_C$ is the volume of the unit cell and $\theta_B$ is the Bragg angle.  $x$ is the depth in the sample that is scaled by $A$. 

The real and imaginary parts of the susceptibility $X$ = $X_1+iX_2$ are given by a set of differential equations 

\begin{equation}
    \frac{dX_1}{dA}=k\left(X_1^2-X_2^2+1\right)+2X_2\left(X_1-y\right)-2gX_1
\end{equation}

\begin{equation}
    \frac{dX_2}{dA}=-\left(X_1^2-X_2^2+1\right)+2X_1\left(X_2k+y\right)-2gX_2. 
\end{equation}

Here,
\begin{equation}
    g=-\frac{f''(0)}{f'(\psi)},\qquad k=\frac{f''(\psi)}{f'(\psi)}.
\end{equation}

For the specific case of symmetric Bragg reflection, the parameter $y$, which defines the deviation of the crystal away from Bragg condition, is given by

\begin{equation}
    y(A)=\frac{\pi V_c\sin(2\theta_B)}{\lambda^2 r_e f'(\psi)}
\left[\Delta\theta+\epsilon(A)\tan\theta_B\right]
-\frac{f'(0)}{f'(\psi)}.
\end{equation}

Here, $\epsilon (A)$ defines the depth dependent strain. In this work, $\epsilon (A)$ is the axial strain along the direction perpendicular to the surface of the crystal defined by the longitudinal strain wave. 

$\Delta\theta$, which defines the difference between the incident angle $\theta$ and the Bragg angle $\theta_B$, i.e.,
\begin{equation}
\Delta\theta = \theta- \theta_{B}.  
\end{equation}

For any $\theta$, the coupled differential equations,  Eq.(4) and Eq.(5), are iteratively solved through the depth of the sample beginning from the depth in the sample where both the real and imaginary parts of the susceptibility are 0. Therefore, we begin the iteration either from  the extinction depth of the sample or from the back surface of the sample. Finally,  we can compute the x-ray reflectivity from the complex susceptibility at the front surface, $X(0)$, by

\begin{equation}
    R=\left|X(0)\right|^2
\end{equation}

In Fig 1, we describe the experimental geometry we use in work and use the formalism by Klar \& Rustichelli \cite{klar1973dynamical} to establish the scaling of \(\Delta \theta\) and \(\omega\) defined by Eq. (2) . The time dependent strain used in the simulation is a longitudinal strain wave  within the (100) plane in silicon with a frequency distribution shown in Fig 2 (d). As a funciton of time, we record the x-ray reflectivity $R(0)$ at every angle $\Delta\theta$ away from the (400) diffraction peak. At each $\Delta\theta$ we take a Fourier transform of the reflectivity.  The position of the peak of the frequency spectrum at every position of $\Delta\theta$ is plotted in Fig 1 (b). 

As it can be seen from Fig 1 (b), the relationship between $\Delta\theta$   and the peak frequency of the oscillation exactly follows the linearity in Eq. (1), as established by Lindenberg \textit{et al.} \cite{lindenberg2000time}.

Finally, in Fig 2, for strain waves with specific frequency distributions, we compute the frequency spectrum of the reflectivity away from all the angles away from the Bragg peak. We then overlay the frequency spectrum of the phonons in the propagating strain wave with the weight of intensity oscillations . We see, from Fig 2, that the weights of the oscillations exactly overlap the phonon frequency spectrum for different phonon spectra, verifying that the amplitude of intensity oscillations scale with the phonon population.

\begin{figure}
\begin{center}
\includegraphics[scale= 0.85]{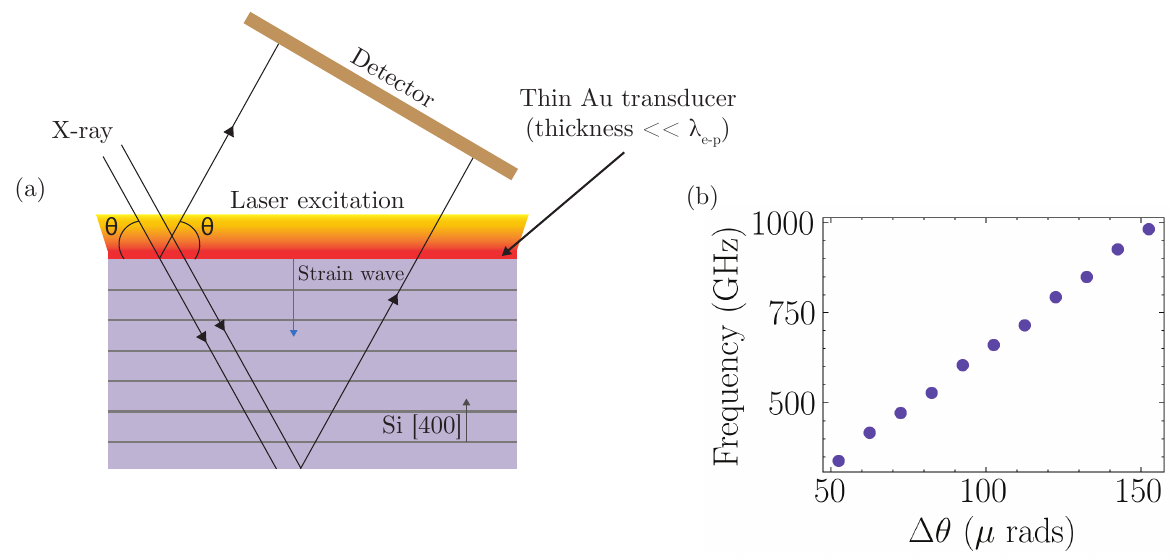}
\caption{\label{fig:epsart} (a) The schematic of the experimental geometry used in this paper. The sample is a (100) silicon deposited with a gold transducer with a thickness much smaller than the characteristic electron- phonon coupling length scale in gold ($\lambda_{e-p}$). A 800 nm femtosceond laser excitation launches a strain wave along the plane normal [100]. A sheet x-ray beam is incident at an angle $\theta$ near the Bragg angle $\theta_B$ of the [400] diffraction plane.  The scattered light at the scattering angle $2\theta$ is collected using an objective lens with the real space image obtained in a detector placed in the imaging condition.  (b) The frequency corresponding to the maximum weight of the frequency spectrum of x-ray reflectivity recorded at different positions ($\Delta\theta$) away from the Bragg peak. The strain used to simulate the time-dependent reflectivity is shown in Fig 2c.}
\end{center}
\end{figure}

\begin{figure}
\begin{center}
\includegraphics[scale= 0.85]{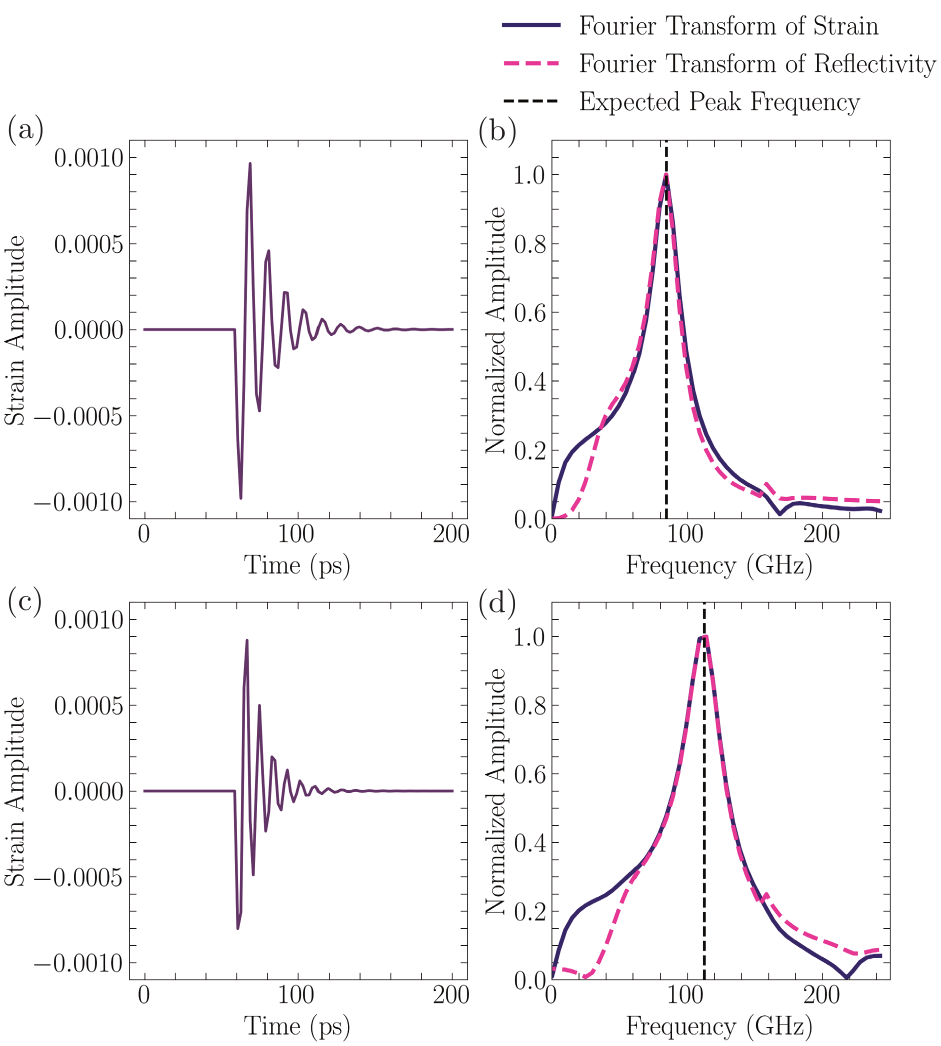}
\caption{\label{fig:epsart} The comparison of the frequency spectra of the applied strain to the frequency spectra of the reflectivity. (a) shows the time-domain behavior of a strain wave whose frequency spectra is plotted in (b). (c) shows the time-domain behavior of a strain wave whose frequency spectra is plotted in (d)}
\end{center}
\end{figure}

 Fig 1 (b) and Fig 2 help us establish  the scaling of \(\omega\) and \(\Delta \theta\) and the scaling of the amplitude of intensity oscillations with the population of phonons respectively. Therefore, for sections 4 and 6,  we use these scalings to calculate intensities and their oscillation rather than using the entire formalism by Klar \& Rustichelli \cite{klar1973dynamical}. This approach greatly shortens the computational time needed to simulate DFXM images in Section 6. 

\section{Spatial resolution of the scattering from coherent phonons}

In any DFXM experiment, the distance between the sample, the objective lens and the detector are optimized such that the image formed in the detector is a real-space resolved image of the sample. The formation of the real-space resolved image requires meeting the imaging condition \cite{simons2017simulating}. In this paper, we assume that the placement sample, the objective lens and the detector are optimized such that the imaging condition is met for the symmetric Bragg geometry. We also assume, the entire sample thickness is within the depth of field of the objective lens.   In the case of dynamical diffraction, however, simply meeting the imaging condition does not guarantee the formation of a real-space image in which there is a linear, one-to-one correspondence between a spatial point in the sample to a spatial point in the detector. 

In applying the Takagi--Taupin equations, one performs the iterative solving of the differential equations (2) and (3) through either the entire sample (if the sample thickness is smaller than the extinction depth of the X-ray) or the thickness that corresponds to the X-ray penetration depth. The iteration accounts for multiple scatterings that could occur within the sample in the case of a dynamical diffraction. Therefore, for a purely dynamical diffraction case, one cannot assign a real space resolution to one point of scattering in the sample when imaging. This is unlike the scattering in the kinematic limit, where scattering arises from a single point in the sample and we can assign a real space resolution in the image plane.

Even though for an unstrained sample, dynamical diffraction results in a signal that can be attributed to the entire sample thickness, for the case of time-dependent intensity oscillations resulting from coherent phonons, it is reasonable to expect that the scatterings resulting in the oscillations originate from localized regions of the sample which correspond to the spatial location of the coherent phonons wavepacket in the sample. 

In order to test the above hypothesis, we perform Takagi--Taupin simulations for coherent phonons of specific spatial extents (number of full cycles). We divide the Takagi--Taupin contributions into three thicknesses: 
\begin{itemize}
    \item Region I is the part of the sample where the wave is already past before we begin our sampling of intensity; 
    \item Region II is the part of the sample that includes everything between the trailing end  of the wave when we begin the sampling and leading end of the wave when we complete the sampling. 
    \item Region III defines the region of the sample that the strain wave does not reach until after the sampling. 
\end{itemize}

We perform the Takagi--Taupin simulations (Equations 3-9) for increasing sample thicknesses as seen in Fig. 3. In Fig 3a, where the strain wave begins from the surface, there is no Region I. We observe the contribution to the frequency oscillations almost immediately for any sample thickness away from the surface. The contribution keeps evolving until a distance of $\sim1600$ nm (end of Region II) and stops evolving for samples thicker. In Fig 3b, we see, when the strain wave begins 200 nm deeper into the sample, i.e. Region I is 200 nm, there is no significant contribution to the time-dependent reflectivity till the sample thickness considered is within Region I. The contribution to the time-dependent reflectivity keeps evolving until we reach a thickness that represents the end of Region II, i.e., $\sim1800$ nm in Fig 3b, and stops evolving for thickness beyond it. 

Therefore, in a DFXM measurement of a phonon sideband, it is possible to assign a real space resolution. The spatial resolution is exactly the spatial extent of the strain wave during the duration of the sampling period, which in turn is determined by the lowest frequency we wish to reconstruct during an experiment.

\begin{figure}
\begin{center}
\includegraphics[scale= 0.4]{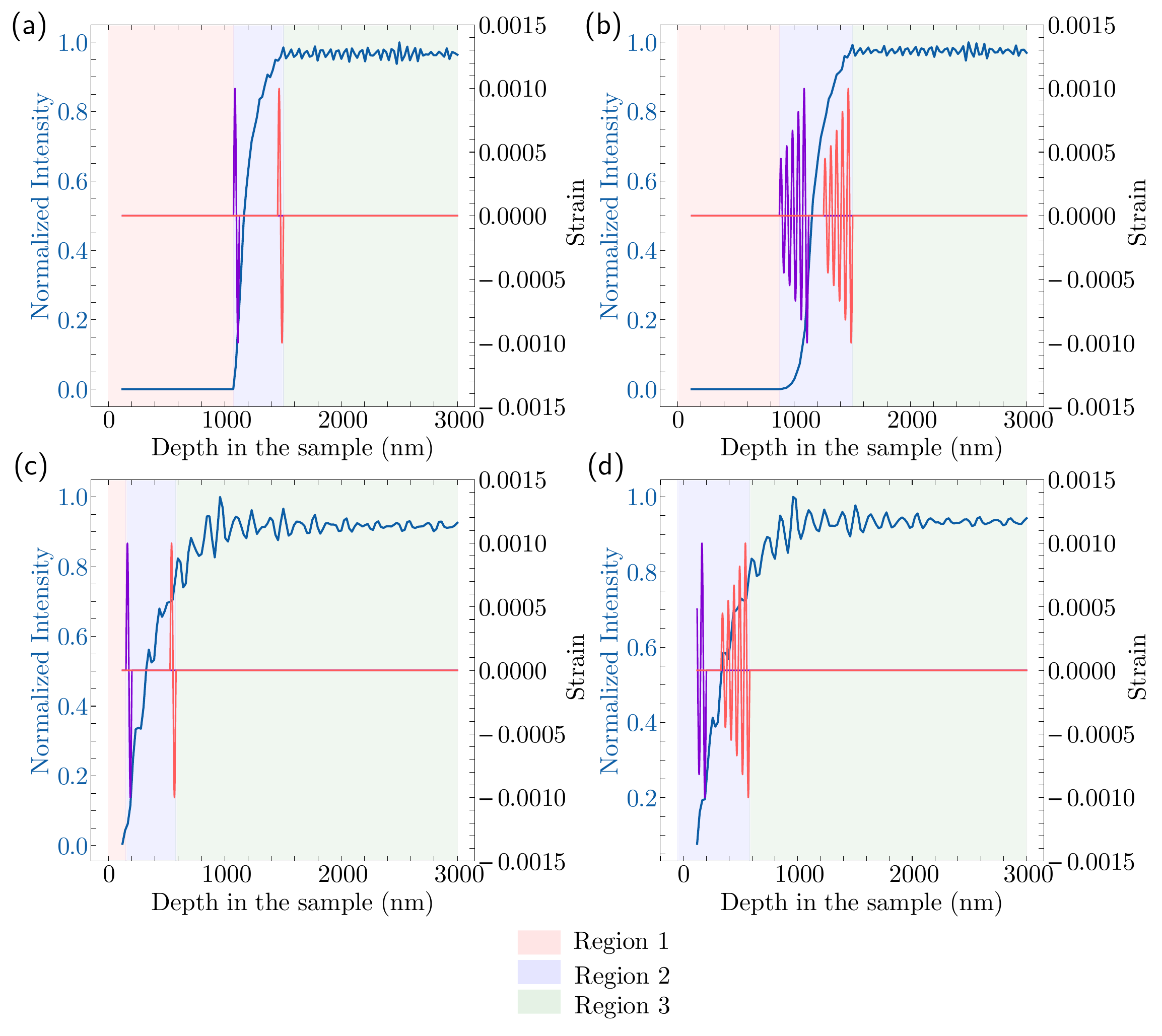}
\caption{\label{fig:wide}Magnitude of intensity oscillations as a function of changing sample thickness. The Takagi-Taupin simulations resulting in the intensity oscillations are performed for strain waves with a primary phonon wavelength of 50 nm in Si. (a) and (c) shows cases where the wave undergoes a 1/e decrease in amplitude over one full cycle while (b) and (d) show cases where the wave undergoes a 1/e decrease in amplitude over five full cycles. (a) and (b) show the cases where we begin sampling the oscillations when the rightmost front of the strain waves is 1125 nm away from the surface of the sample. (c) and (d) show the cases where we begin sampling the oscillations when the rightmost front of the strain waves is 200 nm away from the surface of the sample. The red, blue and green shades in the plots show Region I, II and III discussed in Section 3 respectively. }
\end{center}
\end{figure}

\section{The reciprocal space resolution and its contribution to the damping of time-dependent oscillations}

In general, the reciprocal-space resolution of a DFXM measurement is defined by the bandwidth of the X-ray used, the incoming beam divergences, and the acceptance angle of the objective lens. For the case of Bragg diffraction using kinematic diffraction theory, the reciprocal space resolution for DFXM was first described by Poulsen \textit{et al.} \cite{poulsen2017x}. In Poulsen \textit{et al.} \cite{poulsen2017x}, and the computational framework following the work \cite{poulsen2021geometrical}, the relative intensity of a scattering away from primary Bragg condition is computed assuming a primarily Gaussian distribution  of incoming and out-going x-ray divergences, x-ray energy bandwidth and the acceptance angles of the objective lens. The formalism still assumes that the meeting of Bragg condition is a requirement for observing the scattering. Therefore, the formalism simply computes the Bragg scattering efficiencies for individual sample and objective lens orientation given the distributions in beam divergences and energy bandwidth.  However, for the measurements of the sidebands described in the current work, the reciprocal-space filtering is not directly derivable from Poulsen \textit{et al.} \cite{poulsen2017x}. Therefore, for the specific case of symmetric Bragg geometry, we derive the reciprocal space resolution function in this work. 

For the specific case of imaging coherent phonons described with dynamical diffraction, the scattering itself is dependent upon the experimental geometry (either Bragg or Laue geometry) and asymmetry \cite{lindenberg2000time}. Therefore, developing a general reciprocal space resolution is challenging and beyond the scope of this work. However, we develop the reciprocal space resolution for the specific case of symmetric Bragg reflection geometry in this paper. Such description is in general sufficient because in usual experimental setups, one can apply symmetric Bragg geometries to study coherent acoustic phonons.

Since the objective lens can collect beams scattered at various angles away from the Bragg peak, phonons of multiple frequencies can simultaneous contribute to oscillation signals in DFXM. This is possible because the required angular deviations \(\Delta \theta\) for different phonon frequencies arise from the combined spreads in beam divergence and energy. We aim to determine the contributions of phonons across these frequencies based on the reciprocal space resolution of the measurement.

We first define the nominal parameters for the incoming beam with a central energy \(E_0\) and a bandwidth defined by a top-hat function \(\Delta E/E\). The incoming beam has divergence with the intensity falling off as a Gaussian function away from the center with a full-width-at-half-maximum (FWHM) of $\sigma_i$. \(\theta_{B0}\) defines the Bragg angle for the plane of interest $G$, and the objective lens is centered at an angle \(\Delta \theta_0\) away from the Bragg peak \(\theta_{B0}\).  This configuration optimizes the the collection of intensity oscillations from phonons with frequencies given by $2\pi \nu_0 = v|G|\Delta \theta_0 \cot \theta_B$. 

\begin{figure}
\begin{center}
\includegraphics[scale= 0.65]{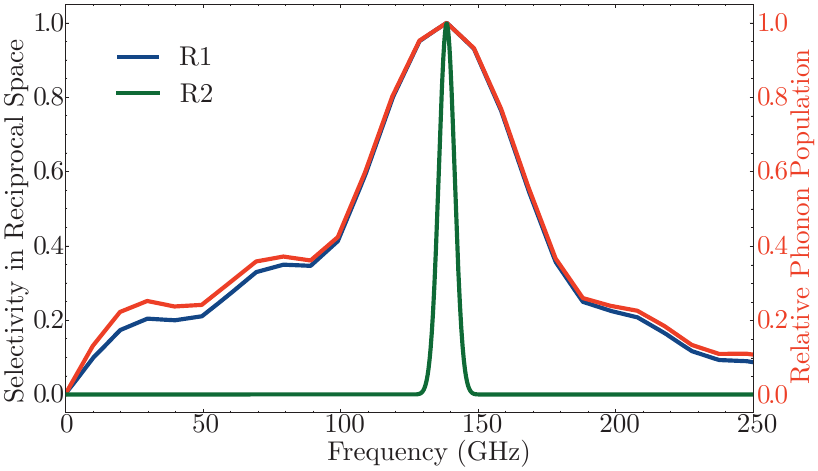}
\caption{\label{fig:wide}Comparison of the frequency spectrum of the populated phonons (solid red) in the material to the DFXM measurable frequency spectrum determined by the reciprocal space resolutions described in Section 4. R1 (solid blue) shows the frequency spectrum of the scattering side bands when using a energy bandwidth of $10^{-4}$ and an incoming beam divergence of $10^{-4}$ rad. R2 (solid green) shows the frequency spectrum of the reflectivity when using a energy bandwidth of $6 \times10^{-6}$ and an incoming beam divergence of $6 \times10^{-6}$ rad.  }
\end{center}
\end{figure}

For developing the reciprocal space resolution, we begin by selecting different positions for the objective lens's acceptance, denoted as \(\Delta \theta_C\). For each acceptance position, we consider a range of energies offset from the central energy. For every chosen energy, we find the corresponding Bragg angle, \(\theta_{B1}\), and calculate the associated angular shift \(\Delta \theta_E\) using \(\Delta \theta_E = \Delta \theta_C - (\theta_{B0} - \theta_{B1})\). We then compute the energy availability \(P_1\) using the top-hat energy distribution, followed by evaluating the angle availability \(P_2\) with a Gaussian distribution of \(\Delta \theta_0 - \Delta \theta_E\). Next, we determine the availability of the specific frequency \(P_3\), where the frequency is defined as \(2 \pi \nu = v G \Delta \theta_E \cot(\theta_{B1})\). We also calculate the Gaussian acceptance \(P_4\) for the CRL based on \(\Delta \theta_0 - \Delta \theta_E\). Finally, we obtain the reciprocal space resolution contribution at each frequency as \(P = P_1 \times P_2 \times P_3 \times P_4\), which gives the weight added for each frequency. 

In Fig 4, we plot the the frequency spectrum of a strain wave with a primary wavelength of 100 nm  whose amplitude undergoes a 1/e decay after three cycles in Si. We compare the frequency selectivity of the phonon frequency spectrum when using a energy bandwidth of $10^{-4}$ and the incoming beam divergence of $10^{-4}$ rad, typical of a DFXM experiment in Fig 4a. In 4b, we use the same phonon distribution but use an energy bandwidth of $6 \times 10^{-6}$ and a vertical beam divergence of $6 \times 10^{-6}$ rad, which can be obtained using the (16 0 0) peak of a Si monochromator \cite{als2011elements}.  The resulting frequency spectrum of the diffuse scattering signal, determined by the reciprocal space resolution, are plotted in Fig 4a and Fig 4b for comparison. 

For a spectrum of the sideband described by a Gaussian, the damping time of the intensity oscillations can be approximated using the FWHM of the spectrum by
\begin{equation}
  \tau_{\text{damping}}\approx 1/{\pi\times\text{FWHM} }.   
\end{equation}
As it can be seen from Fig 4, the FWHM of the reflectivity spectrum when using a x-ray bandwidth of $10^{-4}$  and an incoming beam divergence of $10^{-4}$ rad for the specific phonon spectrum is $\sim100$ GHz, resulting in a damping time of the oscillation $\sim 3$ ps. Our past experiments utilized similar reciprocal space parameters \cite{irvine2025dark}.  Therefore, we do not expect the observation of frequency dependent oscillations in our past experiments. On the other hand, when using a bandwidth of $6 \times 10^{-6}$  and a divergence of $6 \times 10^{-6}$ rad, the (FWHM) is $\sim10$ GHz, which means the intensity oscillations can be observed up to $\sim 30$ ps. 

\section{Optimizing the Phonon Bandwidth for Frequency Selectivity and Long-Lived Intensity Oscillations}



In the Gusev or Thomsen model \cite{chalise2025formalism}, which describes the strain wave generated in thick transducers, the initial distance at which a laser excitation deposits its energy to the lattice is given by the larger of the two numbers: the optical penetration depth of the laser, \(\lambda_p\), or the electron diffusion length, \(\lambda_{e-p}\), over which the excited electrons deposit heat in the lattice. The heat deposited in the metal results in a gradient over this length, causing the phonon wavelength resulting from the thermal expansion to be a distribution \cite{chalise2025formalism}.

However, if the electron–phonon coupling distance is much greater than the film thickness of the metal transducer, the temperature of the entire film is rather close to being isothermal rather than having a gradient in distribution. In such a case, the isothermal expansion of the metal film launches a strain wave in the sample with a wavelength that is exactly double the thickness of the metal film. If the metal film has thickness \(t\), the wave consists of a full cycle: a positive half originating from the part of the wave that gets transmitted into the sample, and a negative half originating from the part that gets reflected at the metal–free surface and then gets reflected back into the metal. The decay of the sinusoid is determined by the acoustic mismatch, and the amplitude in each cycle, $n$, scales by \((1 - R)^n\), where \(R\) is the reflection coefficient of the acoustic wave from the metal–sample (semiconductor) interface, given by
\begin{equation}
R = \left( \frac{Z_1 - Z_2}{Z_1 + Z_2} \right)^2.
\end{equation}
and the transmission coefficient is given by
\begin{equation}
T = \frac{4 Z_1 Z_2}{(Z_1 + Z_2)^2}.
\end{equation}
where \(Z = C_m \rho_m\) is the acoustic impedance of a material, with \(C_m\) the speed of sound and \(\rho_m\) the density.

To verify this prediction, we apply the udkm1Dsim toolbox \cite{schick2014udkm1dsim, schick2021udkm1dsim}. Fig 5 includes the comparisons of strain waves created in a metal (gold in this example) with a weak electron–phonon coupling constant that is then transmitted into the sample (in this case, a silicon single crystal). The electron–phonon coupling results in a characteristic electron diffusion length over which the heat is deposited in the lattice. This length is given by
\begin{equation}
\lambda_{e-p} = \sqrt{\frac{k}{g}}
\end{equation}
where \(k\) is the electron thermal conductivity of the metal film and \(g\) is the electron–phonon coupling constant. For gold, in this example, \(k = 400~\text{W/mK}\) and \(g = 2 \times 10^{16}~\text{J/sm}^3\text{K}\) and therefore, the characteristic length is 150 nm. 

\begin{figure}
\begin{center}
\includegraphics[scale= 0.8]{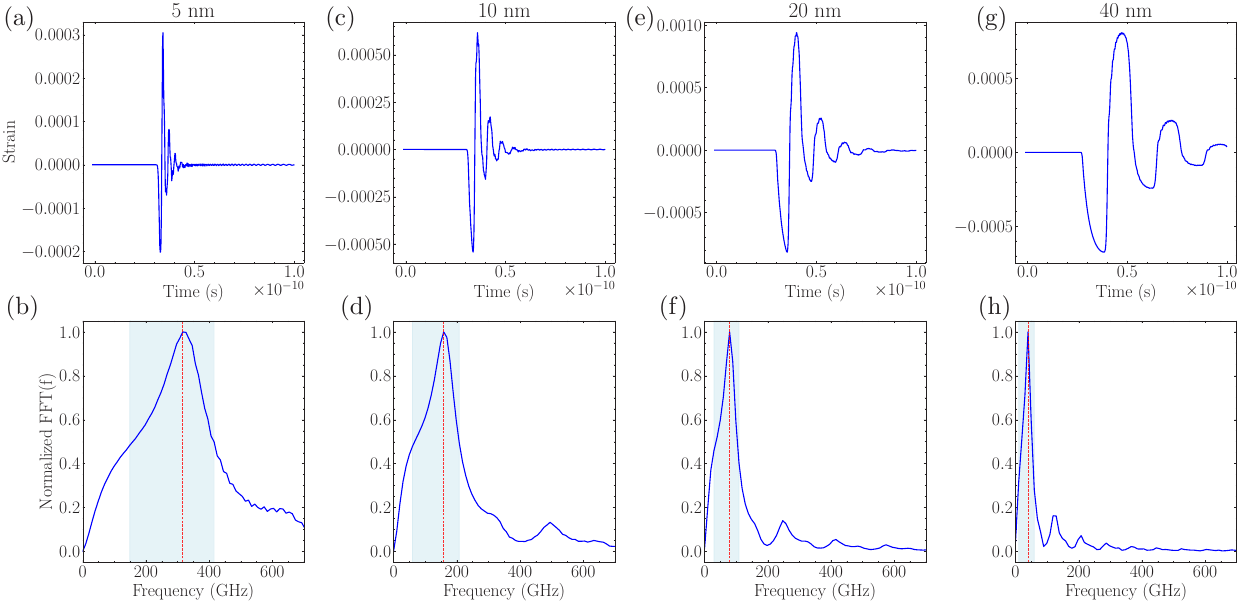}
\caption{\label{fig:wide}Time and frequency behavior of the strain wave in a Si sample generated by ultrafast excitation of a gold transducer of different film thicknesses.The strain wave is simulated using udkm1dsim toolbox. (a) and (b) show the respective time and frequency behavior for a 5 nm gold transducer, (c) and (d) show the behavior for a 10 nm transducer, (e) and (f) for a 20 nm transducer and (g) and (h) for a 40 nm transducer. The red lines in the frequency-domain plots represent the expected frequency based on the theory described in Section 5 and the sky-blue transparent region represents the FWHM for each spectrum. }
\end{center}
\end{figure}

For gold films with thicknesses much smaller than the electron–phonon coupling length, the strain wave in silicon takes the form of a sinusoidal wave whose amplitude in each cycle scales by \((1 - R)^n\).  As it can be seen from Fig 5, the frequency of the strain wave is centered at the value of
\begin{equation}
  \nu_{0} = \frac{c_{\text{gold}}}{2 \times (\text{gold film thickness})}
\end{equation} with $ c_\text{gold}$ representing the speed of sound in gold.  
This is expected from the discussion in the paragraph above, i.e., the whole film undergoes an isothermal expansion.

As the film thickness increases and becomes comparable to the electron–phonon coupling length, the film starts experiencing a temperature gradient, and the strain wave starts approaching the profile given by the Gusev model \cite{holstad2022x}.

As compared to the Gusev model, the relative width of the spectra is smaller when the film thickness of the transducer is much smaller than the electron-phonon coupling distance. Nonetheless, the spectrum still has a width because the decay of the wave at each cycle. For samples in which the acoustic mismatch is higher, this decay is smaller, and therefore more number of intensity oscillations can be recorded before the signal of the oscillation dampens. However, selecting sample combinations with high acoustic mismatch is problematic for real space resolution, as a large acoustic mismatch results in a wave with a larger spatial extent. As it can be seen from Section 3, the large spatial extent results in a large spatial resolution, which is not ideal for an experiment. Alternatively, as described in Section 4, one can instead improve the reciprocal space resolution of the experiment to optimize the lifetime of oscillations, without compromising the real-space resolution. 

\section{Simulations of the imaging of coherent phonons propagation using DFXM}

\begin{figure}
\begin{center}
\includegraphics[scale= 0.8]{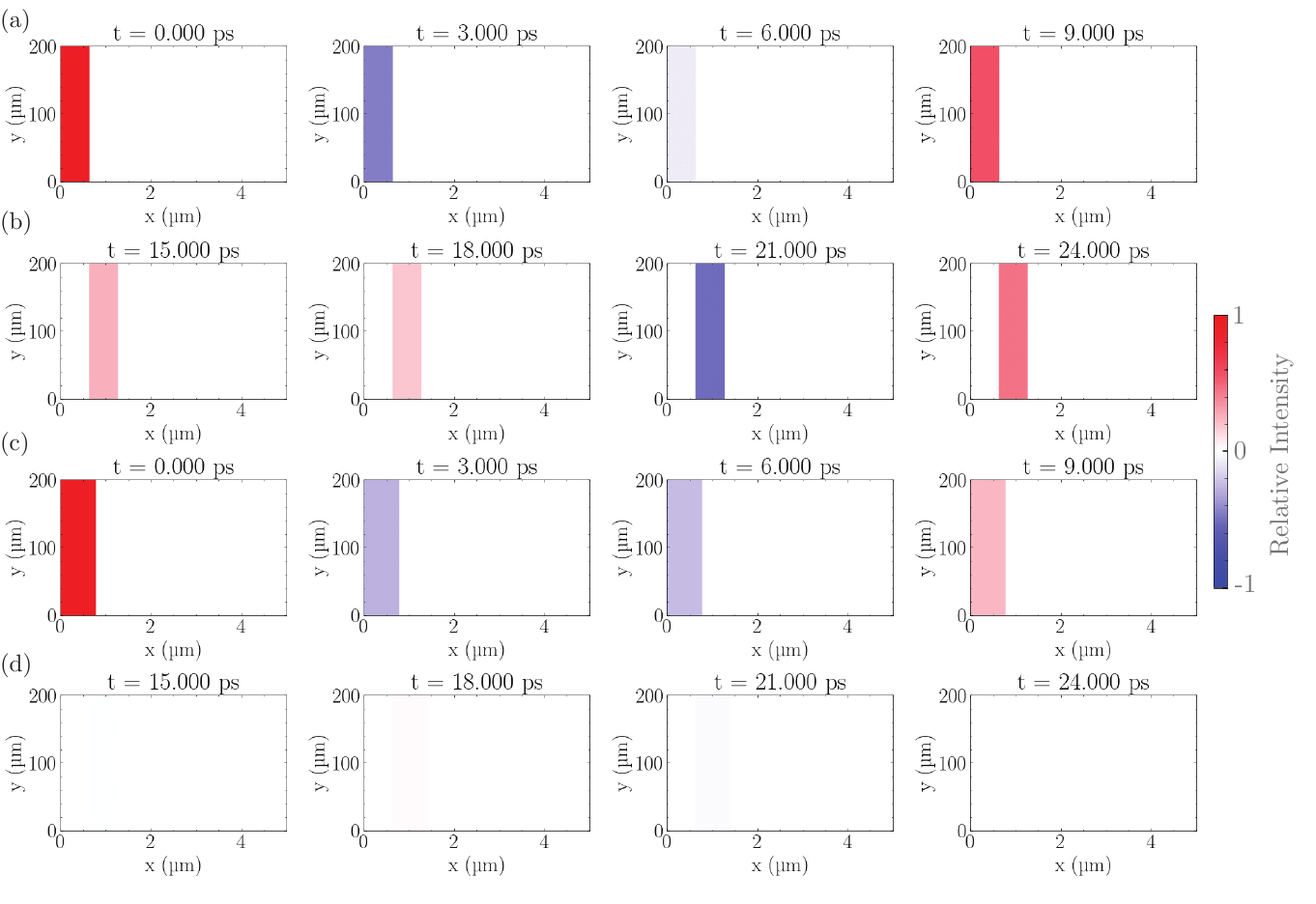}
\caption{\label{fig:wide}Simulated DC-filtered DFXM images of intensity oscillations due to coherent phonons at different sampling intervals and reciprocal space resolution. (a) shows the intensity oscillations in the sampling interval of 0-9 ps with energy bandwidth and beam divergence of $6\times 10^{-6}$.  (b) shows the intensity oscillation in the sampling interval of 15-24 ps with an energy bandwidth and a beam divergence of $6\times 10^{-6}$. (c) and (d) show the intensity oscillations in the same sampling duration as (a) and (b) respectively but with an an energy bandwidth and a beam divergence of $10^{-4}$. }
\end{center}
\end{figure}

In Section 2-4 we use the Takagi-Taupin equations for the case of symmetric Bragg reflections to establish that the x-ray reflectivity shows frequency dependent oscillations at specific angles away from the Bragg peak. We also show the amplitude  of the oscillation at a particular frequency scales with the phonon population at that frequency in a spectrum. Finally, we have established that the spatial origin of the oscillation within the sample lie within the spatial extent of the strain wave in the sample. Additionally, in Section 5, we have established using the udkm1Dsim toolbox that if the thickness of the transducer is much smaller than the electron-phonon coupling length, we produce coherent phonons in the sample whose decay in every cycle is determined by the acoustic mismatch between the transducer and the sample. Using these established facts, along with the reciprocal space resolution function developed in Section 4, we can create a simulation model that can be used to forward-model the time-dependent oscillation of the reflectivity signal imaged in a DFXM experiment. We develop such a simulation in this section.

Our simulation uses the thickness of the transducer, the power and area of the incident laser, the bandwidth and the vertical divergence of the x-ray pulse as the input parameters. We use the initial amplitude determined from our udkm1Dsim toolbox as the scaling factor which scales with the power per unit area to determine the amplitude of the strain wave that is created. In addition, we use the velocity of sound and the density of the transducer and the sample to determine the acoustic mismatch and therefore the amplitude of the strain wave in the sample.

We use the Fourier transform of the strain wave and the reciprocal space resolution function determined by the beam divergence and the x-ray bandwidth to select the individual frequencies of the reflectivity that are imaged during the experiment. As a function of time, we forward-model the image with intensity oscillations at frequencies and their weights selected by the reciprocal space resolution. 

Finally, we use the x-ray beam energy and the lattice spacing of the imaged plane in the sample to predict the scattering angle and therefore model the projection of the image on the detector. To assign the real-space resolution, we use the results from Section 3, where the spatial resolution in the region showing intensity oscillations corresponds to the full spatial extent of the wave within the sampling period. 

We note that the image generated by our simulations are DC-filtered images which only includes the contribution from the side-bands and not the primary Bragg diffraction. Similarly, at any given time, the intensity detected at a particular angle, 
$\Delta\theta$, may cover a smaller spatial extent than that in our simulations. However, to fully sample the lowest-frequency oscillations during post-processing, the user must sum the intensities from multiple pixels to capture the complete oscillation pattern. Thus, the smallest detector region that must be integrated to observe the lowest-frequency oscillations defines the spatial resolution in our simulations.

 Fig. 6 (a and c) shows the simulations of the time evolution of the intensity oscillations as DFXM signals in Si with a 20 nm gold transducer with a bandwidth of \(10^{-6}\) and a vertical divergence of \(10^{-6}\) radians. Due to the narrow reciprocal space resolution, the range of frequencies selected is small and consequently the dephasing of the signal occurs slowly.

Fig. 6 (b and d) show the time evolution of intensity oscillation when the bandwidth and the vertical divergence are $ 10^{-4}$ and $10^{-4}$ radians respectively. The wider reciprocal space resolution results in a faster decoherence due to larger bandwdith of phonon frequencies selected. 

As illustrated in Fig 6, it is imperative for DFXM experiments to maintain a narrow reciprocal space resolution to ensure the intensity oscillations are longer lived and can be studied deep within the sample. 

\section{Discussions and Conclusions}

In this work, we develop a formalism to simulate DFXM images of intensity oscillations away from the Bragg peak induced by coherent phonons. For developing the simulations, we first establish the relationship between the frequency distribution of phonons to the frequency distribution of the intensity oscillation and their position in reciprocal space, using Takagi-Taupin equations. We then establish the expected real space resolution observed for the intensity oscillation and finally build the reciprocal space resolution computation that enables developing the simulation.

The developed formalism addresses the experimental observations from dynamical diffraction for measurements of coherent phonon propagation, which is not adequately addressed with kinematic theory. The approach of observing the intensity oscillations away from the Bragg peak also enables studying high-frequency phonons, whose frequency reconstruction is fundamentally limited by spatial resolution when simply measuring the shift in Bragg peak \cite{chalise2025formalism}.

The developed reciprocal space resolution for the symmetric Bragg geometry also provides the requirements on the beam parameters for long-lived observations of the oscillations. For studying phonon damping and interactions deep within the sample, it is imperative to maintain a narrow bandwidth and beam divergence. To maintain sufficient x-ray intensity while maintaining such narrow bandwidth and divergence is a challenge for present-day XFEL technology. However, with future upgrades, such possibilities of high brilliance can be possible.

We note that the paper doesn't explicitly discuss about absolute intensities and the signal-to-noise ratios of the observed oscillations when developing the formalism. We plan to address them in the future directly through analysis of experimental results. We also note, the formalism is specific to scattering sidebands from coherent phonons.  Diffuse scattering from incoherent phonons provide means for thermometry and thermal transport measurements \cite{chalise2022temperature,chalise2025formalism}. In our future works, we also plan to address DFXM imaging of diffuse scattering from incoherent phonons experimentally and theoretically.

\begin{acknowledgments}

D.C., B.K., A.M.L. and L.D.M. acknowledge the support by the Department of Energy, Office of Science, Basic Energy Sciences, Materials Sciences and Engineering Division, under Contract DEAC02-76SF00515. D.P.L. acknowledges the support by Stanford Graduate Fellowship. 

\end{acknowledgments}
\nocite{*}

\bibliography{bib}

\end{document}